\documentclass[11pt]{article}
\usepackage[letterpaper,margin=1.00in]{geometry}
\usepackage[T1]{fontenc}
\usepackage{lmodern}
\usepackage{amsmath,amssymb,mathtools,bm}
\usepackage{booktabs,array,tabularx,graphicx,microtype}
\usepackage[square,numbers,sort&compress]{natbib}
\usepackage{xcolor}
\definecolor{linkblue}{RGB}{0,50,232}
\definecolor{equationpink}{RGB}{255,0,127}
\usepackage[colorlinks=true,linkcolor=linkblue,citecolor=linkblue,urlcolor=linkblue]{hyperref}
\makeatletter
\renewcommand{\eqref}[1]{\textup{\hyperref[#1]{\textcolor{equationpink}{\tagform@{\ref*{#1}}}}}}
\makeatother
\numberwithin{equation}{section}
\newcommand{\dd}{\mathrm d}
\newcommand{\GB}{\mathcal L_{\rm GB}}
\newcommand{\E}{\widehat{\mathcal E}}
\newcommand{\doilink}[1]{\href{https://doi.org/#1}{\textcolor{linkblue}{doi:#1}}}

\title{\bfseries Exact Black Branes in General Dimensions in\\
Higher-Curvature Scalar--Tensor Gravity}
\author{Tianhao Wu\thanks{Corresponding author: twu49@illinois.edu}\\
\normalsize Department of Physics, University of Illinois Urbana-Champaign,\\[-2pt]
\normalsize Urbana, Illinois 61801, USA}
\date{}

\begin{document}
\maketitle

\begin{abstract}
We obtain two families of neutral planar black branes in a potential-free dilaton-coupled Lovelock--Horndeski theory. Heterotic compactification relates higher-dimensional curvature to scalar dynamics and motivates the interaction structure. At fixed scalar zero mode, the horizon radius and temperature vary while the complete stationary Wald entropy remains constant and positive. The exponential scalar factor at the horizon exactly compensates the area growth, giving this behavior a direct geometric explanation. An exact coefficient analysis determines the two four-dimensional branches and their continuation to every higher dimension through closed coupling relations. The linear family is conformally flat and leaves the AdS scale free at fixed couplings; the fractional family is Weyl curved and fixes that scale. The common kinetic relation also excludes their realization through a real one-scalar flat-torus reduction of a metric--dilaton parent.
\end{abstract}


\clearpage

\section{Introduction}

Black-hole entropy relates the geometry of a horizon to the dynamics of the gravitational field. In scalar--tensor gravity, the scalar can participate both in supporting the geometry and in determining its entropy. Lovelock--Horndeski gravity places these two roles within a common geometric construction: higher-dimensional curvature generates scalar interactions under dimensional reduction, relating the scalar and metric sectors at the level of the action \cite{VanAcoleyenVanDoorsselaere2011,CharmousisGouterauxKiritsis2012}.
Exact black branes make it possible to follow that relation through the nonlinear field equations and into a horizon observable.

The theoretical basis combines the Lovelock organization of curvature with the second-order scalar dynamics of Horndeski theory and
covariant Galileons \cite{Lovelock1971,Horndeski1974,NicolisRattazziTrincherini2009,
DeffayetEspositoFareseVikman2009}. Heterotic string compactification provides a physical realization of their geometric connection \cite{WuStone2026}. In particular, Ref.~\cite{WuStone2026} derives a Lovelock--Horndeski branch from the metric--dilaton effective action and establishes its holographic anomaly and $a$-theorem. The corresponding holographic renormalization yields finite boundary responses and Ward identities \cite{WuHolographicRenormalization2026}. The curvature--dilaton coefficients and their allowed field redefinitions are central to this connection with string effective theory
\cite{Zwiebach1985,TseytlinAmbiguity1986,MetsaevTseytlin1987,GrossSloan1987}.

Exact Lovelock black holes and dilatonic higher-curvature solutions already show how curvature and scalar fields can change the relation between a horizon and its asymptotic region \cite{BoulwareDeser1985,Wheeler1986,CrisostomoTroncosoZanelli2000,Cai2002,CallanMyersPerry1989,KantiEtAl1996,AgurtoSepulveda2023}.
The black branes constructed here reveal a different relation: their horizon radii and temperatures can vary while the complete stationary entropy stays fixed. A scalar weight compensates the area growth along each family. The conditions sustaining this behavior also determine the allowed couplings and constrain a proposed higher-dimensional origin. The entropy mechanism, the exact solution families and the inverse compactification problem can consequently be studied within the same nonlinear system.

We consider the neutral, potential-free theory with an exponential dilaton weight,
\begin{equation}
 S_{\Omega}=\frac{1}{16\pi G_d}\int\dd^dx\sqrt{-g}\,e^{\lambda\phi}
 \Big( R+\alpha_0X+\alpha_1\GB
 +\alpha_2G^{\mu\nu}\phi_\mu\phi_\nu
 +\alpha_3X\Box\phi+\alpha_4X^2\Big),
 \label{eq:action}
\end{equation}
where $X=g^{\mu\nu}\phi_\mu\phi_\nu$ and
\begin{equation}
 \GB=R_{\mu\nu\rho\sigma}R^{\mu\nu\rho\sigma}
 -4R_{\mu\nu}R^{\mu\nu}+R^2.
\end{equation}
The scalar weight keeps the Gauss--Bonnet interaction dynamically active in four dimensions. The Einstein term and scalar derivative
interactions support the AdS curvature together, even though the action contains no scalar potential or bare cosmological constant.
The logarithmic scalar has a finite asymptotic kinetic density, so its role can be followed continuously from the horizon to the asymptotic region. We take $\phi$, $\lambda$ and $\chi$ to be dimensionless; $\alpha_0$ is dimensionless and $\alpha_1,\ldots,\alpha_4$ have dimension $L^2$.

We determine two exact black-brane families and their closed coupling relations in every dimension $d\geq4$. In four dimensions, we classify the nondegenerate solutions with a logarithmic scalar, a single-power deformation of the AdS metric, nonzero couplings and a continuously variable blackening parameter. The two roots have linear and \( 3/2 \) blackening powers. The linear family is conformally flat and retains a free AdS radius at fixed couplings; the fractional family has nonzero Weyl curvature and fixes that radius.
Varying the radial fields before gauge fixing retains all constraints. Explicit coefficient equations determine the branches and reduce to four independent coupling conditions for each family in every dimension.

The complete stationary Wald calculation gives the same positive entropy for the two families at fixed scalar zero mode. This is a property of the full scalar--tensor action: its curvature and derivative interactions are included in the horizon analysis. The scalar weight precisely offsets the horizon-area dependence, and the Einstein variables express the result as a constant transverse area. These solutions thus give a direct analytic account of how scalar hair changes the relation between geometric size, temperature and horizon entropy.

Their higher-dimensional interpretation follows from an inverse problem. The black-brane equations select coupling relations; a specified compactification must reproduce those same relations to realize either family. For the real one-scalar flat-torus reduction in section~\ref{sec:parent}, the common kinetic relation gives an exact no-go independently of the four-derivative frame coefficients. The existence of the black branes therefore supplies a concrete restriction on the compactification data of a proposed parent theory. In the Lovalock-Horndeski theory studied in~\cite{WuCompanionHairy2026}, planar and hyperbolic hairy black holes coexist in one fixed action and obey a covariant charge balance.

\section{Lovelock--Horndeski gravity}
\label{sec:method}

\subsection{Covariant equations}

We write
\begin{equation}
 \Omega=e^{\lambda\phi},\qquad
 X=\nabla_\rho\phi\nabla^\rho\phi,
 \label{eq:covariantdefinitions}
\end{equation}
and normalize the Euler tensors through the bulk variation
\begin{equation}
 \delta S_\Omega=\frac{1}{16\pi G_d}\int\dd^dx\sqrt{-g}
 \left(\mathcal E_{\mu\nu}\delta g^{\mu\nu}
 +\mathcal E_\phi\delta\phi\right).
 \label{eq:covariantvariation}
\end{equation}
The complete metric equation is
\begin{equation}
\begin{aligned}
\frac{\mathcal E_{\mu\nu}}{\Omega}={}&
G_{\mu\nu}+g_{\mu\nu}(\lambda\Box\phi+\lambda^2X)
-\lambda\nabla_\mu\nabla_\nu\phi
-\lambda^2\nabla_\mu\phi\nabla_\nu\phi +\alpha_0\left(
\nabla_\mu\phi\nabla_\nu\phi-\frac12g_{\mu\nu}X\right)\\
&+\alpha_1\Bigl(
\mathcal H^{\rm GB}_{\mu\nu}
-2\Pi_{\mu\rho\sigma\nu}\left(
\lambda\nabla^\rho\nabla^\sigma\phi
+\lambda^2\nabla^\rho\phi\nabla^\sigma\phi\right)\Bigr)\\
&+\alpha_2\Bigl\{
-\frac12R\nabla_\mu\phi\nabla_\nu\phi
+2R_{\rho(\mu}\nabla_{\nu)}\phi\nabla^\rho\phi
+R_{\mu\rho\nu\sigma}\nabla^\rho\phi\nabla^\sigma\phi
-\frac12XG_{\mu\nu}\\
&\qquad
+(\nabla_\mu\nabla_\rho\phi)(\nabla_\nu\nabla^\rho\phi)
-(\Box\phi)\nabla_\mu\nabla_\nu\phi\\
&\qquad
+g_{\mu\nu}\Bigl(
-\frac12(\nabla_\rho\nabla_\sigma\phi)
          (\nabla^\rho\nabla^\sigma\phi)
+\frac12(\Box\phi)^2
-R_{\rho\sigma}\nabla^\rho\phi\nabla^\sigma\phi\Bigr)\\
&\qquad
+\lambda\Bigl(
\nabla_{(\mu}\phi\,
  (\nabla_{\nu)}\nabla_\rho\phi)\nabla^\rho\phi
-\frac12X\nabla_\mu\nabla_\nu\phi
-\frac12(\Box\phi)\nabla_\mu\phi\nabla_\nu\phi\\
&\hspace{5.4em}
+\frac12g_{\mu\nu}\left(
X\Box\phi-\nabla^\rho\phi\nabla^\sigma\phi\,
             \nabla_\rho\nabla_\sigma\phi\right)
\Bigr)  \Bigr\}\\
&+\alpha_3 \left(
(\Box\phi-\lambda X)\nabla_\mu\phi\nabla_\nu\phi
-2\nabla_{(\mu}\phi\,
  (\nabla_{\nu)}\nabla_\rho\phi)\nabla^\rho\phi +g_{\mu\nu}\left(
\nabla^\rho\phi\nabla^\sigma\phi\,
  \nabla_\rho\nabla_\sigma\phi
+\frac{\lambda}{2}X^2\right)\right)\\
&+\alpha_4\left(
2X\nabla_\mu\phi\nabla_\nu\phi
-\frac12g_{\mu\nu}X^2\right)=0.
\end{aligned}
\label{eq:metricEOM}
\end{equation}
The curvature tensors in \eqref{eq:metricEOM} are
\begin{equation}
\begin{aligned}
\Pi^{\mu\nu\rho\sigma}={}&
2R^{\mu\nu\rho\sigma}
-4\left(g^{\mu[\rho}R^{\sigma]\nu}
-g^{\nu[\rho}R^{\sigma]\mu}\right)
+2R g^{\mu[\rho}g^{\sigma]\nu},\\
\mathcal H^{\rm GB}_{\mu\nu}={}&
2\Bigl(RR_{\mu\nu}-2R_{\mu\rho}R_\nu{}^\rho
-2R^{\rho\sigma}R_{\mu\rho\nu\sigma}
+R_\mu{}^{\rho\sigma\lambda}R_{\nu\rho\sigma\lambda}\Bigr)
-\frac12g_{\mu\nu}\GB.
\end{aligned}
\label{eq:covarianttensors}
\end{equation}
The complete scalar equation is
\begin{equation}
\begin{aligned}
\frac{\mathcal E_\phi}{\Omega}={}&
\lambda R-\alpha_0(2\Box\phi+\lambda X)
+\lambda\alpha_1\GB -\alpha_2\Bigl(
2G^{\mu\nu}\nabla_\mu\nabla_\nu\phi
+\lambda G^{\mu\nu}\nabla_\mu\phi\nabla_\nu\phi\Bigr)\\
&+\alpha_3\Bigl\{
2\Bigl( (\nabla_\mu\nabla_\nu\phi)
        (\nabla^\mu\nabla^\nu\phi)
-(\Box\phi)^2
+R_{\mu\nu}\nabla^\mu\phi\nabla^\nu\phi\Bigr)\\
&\qquad
+4\lambda\nabla^\mu\phi\nabla^\nu\phi\,
  \nabla_\mu\nabla_\nu\phi
+\lambda^2X^2\Bigr\}\\
&-\alpha_4\Bigl(
8\nabla^\mu\phi\nabla^\nu\phi\,
  \nabla_\mu\nabla_\nu\phi
+4X\Box\phi+3\lambda X^2\Bigr)=0.
\end{aligned}
\label{eq:scalarEOM}
\end{equation}
Both equations contain derivatives of at most second order and obey
the off-shell identity
\begin{equation}
 2\nabla^\mu\mathcal E_{\mu\nu}
 +\mathcal E_\phi\nabla_\nu\phi=0.
 \label{eq:covariantNoether}
\end{equation}
A constant scalar displacement rescales the action by $e^{\lambda c}$
and preserves its classical field equations. The parameter $r_0$ in the
logarithmic profile fixes this scalar zero mode and hence the action
normalization and horizon entropy. We hold $r_0$ fixed when varying the
blackening modulus within either family.

\subsection{Exact radial reduction}

We set
\begin{equation}
 d=p+2,\qquad p\geq2,
 \label{eq:dimension}
\end{equation}
and begin with four independent radial functions in future-ingoing coordinates,
\begin{equation}
 \dd s_d^2=-N(r)^2F(r)\dd v^2+2N(r)\dd v\dd r
 +\Sigma(r)^2\dd\bm{x}_p^2,\qquad \phi=\phi(r).
 \label{eq:pregauge}
\end{equation}
This ansatz describes the static, homogeneous, and isotropic planar sector in future-ingoing radial gauge, with $\mathcal L_{\partial_v}g_{\mu\nu}=0=\mathcal L_{\partial_v}\phi$.
The chart is regular at a simple zero of $F$. Independent variation of $N$ and $\Sigma$ gives the lapse and transverse constraints.
Write $\dd s_d^2=h_{AB}\dd x^A\dd x^B+\Sigma^2\dd\bm{x}_p^2$ and let $D_A$ denote the covariant derivative of the two-dimensional base.  Define
\begin{equation}
 S_{AB}=D_AD_B\Sigma,\qquad B_\Sigma=D^AD_A\Sigma,\qquad
 U=D_A\Sigma D^A\Sigma,\qquad v_A=D_A\phi,\qquad X=v_Av^A.
 \label{eq:warpedblocks}
\end{equation}
Here $R_h$ denotes the Ricci scalar of $h_{AB}$.  For a flat fiber, the
complete invariants entering the action reduce to
\begin{equation}
\begin{aligned}
R_d={}&R_h-\frac{2pB_\Sigma}{\Sigma}
-\frac{p(p-1)U}{\Sigma^2},\\
\GB={}&-\frac{2p(p-1)}{\Sigma^2}R_hU
+\frac{4p(p-1)}{\Sigma^2}
\left(B_\Sigma^2-S_{AB}S^{AB}\right)\\
&+\frac{4p(p-1)(p-2)}{\Sigma^3}B_\Sigma U
+\frac{p(p-1)(p-2)(p-3)}{\Sigma^4}U^2,\\
Z_\phi\equiv G^{\mu\nu}\phi_\mu\phi_\nu={}&
\frac{p}{\Sigma}\left(B_\Sigma X-S_{AB}v^Av^B\right)
+\frac{p(p-1)}{2\Sigma^2}UX,\\
H_\phi\equiv\Box_d\phi={}&D^AD_A\phi
+\frac p\Sigma D_A\Sigma v^A.
\end{aligned}
\label{eq:warpedreduction}
\end{equation}
Hence the reduced density per unit planar volume is explicitly
\begin{equation}
{\cal L}_{\rm red}=N\Sigma^p e^{\lambda\phi}
\left[R_d+\alpha_0X+\alpha_1\GB+\alpha_2Z_\phi
+\alpha_3XH_\phi+\alpha_4X^2\right].
\label{eq:reducedLagrangian}
\end{equation}
In the static exterior, $\dd v=\dd t+\dd r/(NF)$ brings
\eqref{eq:pregauge} to
$-N^2F\dd t^2+\dd r^2/F+\Sigma^2\dd\bm{x}_p^2$.
We vary \eqref{eq:reducedLagrangian} with respect to $N,F,\Sigma,$ and $\phi$ before fixing $\Sigma=r$ or the lapse. The four radial Euler--Lagrange equations are
\begin{equation}
 \E_Q=\frac{\partial{\cal L}_{\rm red}}{\partial Q}
 -\frac{\dd}{\dd r}\frac{\partial{\cal L}_{\rm red}}{\partial Q'}
 +\frac{\dd^2}{\dd r^2}
 \frac{\partial{\cal L}_{\rm red}}{\partial Q''}=0,
 \qquad Q\in\{N,F,\Sigma,\phi\}.
 \label{eq:Euleroperator}
\end{equation}
They are exactly the projections of the covariant equations:
\begin{equation}
\begin{aligned}
\E_N&=-\frac{2\Sigma^p}{N}\mathcal E_{vr},&
\E_F&=N\Sigma^p\mathcal E_{rr},\\
\E_\Sigma&=-2N\Sigma^{p-3}\delta^{ij}\mathcal E_{ij},&
\E_\phi&=N\Sigma^p\mathcal E_\phi .
\end{aligned}
\label{eq:covariantReducedBridge}
\end{equation}
The areal choice $\Sigma=r$ fixes the radial coordinate.  The exact branches then have constant $N$, which is normalized to one by a constant rescaling of $v$.  The third and fourth derivatives in \eqref{eq:Euleroperator} cancel.  In the diagonal static chart, $\mathcal E_{tr}=0$ identically by symmetry; transforming back to future-ingoing coordinates gives $\mathcal E_{vv}=-NF\mathcal E_{vr}$.  Thus the displayed projections include the remaining metric equation in this static sector, consistently with \eqref{eq:covariantNoether}.
A logarithmic scalar turns the exponential weight into a radial power.
Together with a single-power deformation of the AdS metric, it reduces the field equations to coefficient equations for the blackening exponent, scalar slope and couplings. We take
\begin{equation}
 F(r)=\frac{r^2}{L^2}-\mu r^b,
 \qquad \phi(r)=\chi\log\frac r{r_0},
 \qquad \tau=\lambda\chi,
 \label{eq:seed}
\end{equation}
After variation and gauge fixing, the scalar invariants and weighted volume element are
\begin{equation}
\begin{aligned}
X&=\chi^2\left(\frac1{L^2}-\mu r^{b-2}\right),\\
H_\phi&=\chi\left[\frac{p+1}{L^2}
-(b+p-1)\mu r^{b-2}\right],\\
\sqrt{-g}\,e^{\lambda\phi}&=r^p\left(\frac r{r_0}\right)^\tau.
\end{aligned}
\label{eq:seedkinematics}
\end{equation}
For the four-dimensional classification, each weighted residual is a polynomial of degree at most two in $\mu$,
\begin{equation}
 e^{-\lambda\phi}\E_A
 =\sum_{j=0}^{2}\mu^j{\cal R}^{A}_{j}
 (r;b,\tau;p;\alpha_i),
 \qquad A=N,F,\Sigma,\phi.
 \label{eq:coeffextract}
\end{equation}
We require $\mu$ to vary freely, so each coefficient of this polynomial must vanish. Evaluating these coefficients at the positive reference radius $r=1$ gives eleven necessary algebraic conditions. We solve them in the nondegenerate sector specified below and divide out only nonzero overall factors. We then substitute each candidate into all four original radial equations \eqref{eq:Euleroperator} at arbitrary $r$. For the $E_a$ and $E_b$ exponents obtained below, we also extract every radial coefficient. For $E_b$, we use $r=x^2$, $x>0$, to handle the fractional powers.

A positive simple horizon obeys
\begin{equation}
 r_h^{\,2-b}=\mu L^2,\qquad
 F'(r_h)=\frac{(2-b)r_h}{L^2},\qquad
 T=\frac{(2-b)r_h}{4\pi L^2}.
 \label{eq:horizon-general}
\end{equation}
The nondegenerate sector with nonzero couplings has $L^2>0$, $\mu>0$, $b<2$,
$\lambda\chi\ne0$, and $\prod_{i=0}^4\alpha_i\ne0$. The natural
dimensionless coefficient coordinates are
\begin{equation}
 (u_0,u_1,u_2,u_3,u_4)=\left(
 \alpha_0\chi^2,\frac{\alpha_1}{L^2},
 \frac{\alpha_2\chi^2}{L^2},
 \frac{\alpha_3\chi^3}{L^2},
 \frac{\alpha_4\chi^4}{L^2}\right).
 \label{eq:ucoordinates}
\end{equation}
These dimensionless combinations are independent of the scalar normalization and make the $p$ dependence of the coupling families explicit.

\section{Four-dimensional black branes}
\label{sec:fourD}

Appendix~\ref{app:Ecoefficients} lists the coefficient polynomials $C_1,\ldots,C_{11}$ and identifies the radial equation from which each is obtained. At $p=2$, two exact combinations give
\begin{equation}
C_1+C_3-C_2=(\tau+2)(b-2),\qquad
C_4-C_5=u_0-\tau(\tau-1).
\label{eq:ideal-factors}
\end{equation}
An outer simple horizon requires $b<2$. The first identity therefore fixes $\tau=-2$, and the second then gives $u_0=6$.

On this locus, $C_1=0$ and $C_5-2C_1=0$ determine
\begin{equation}
u_3=-3u_2-24u_1,\qquad
u_4=5u_2+48u_1,\qquad v=u_2+16u_1.
\label{eq:Etriangularcouplings}
\end{equation}
Denote by $\widetilde C_i$ the coefficients after these substitutions.
The transverse equations reduce to
\begin{equation}
\widetilde C_7=-(b-1)(b-2)(v+2),\qquad
\widetilde C_8=-(2b-3)(b-2)v.
\label{eq:Etriangulartransverse}
\end{equation}
Eliminating $v$ gives the polynomial identity
\begin{equation}
(2b-3)\widetilde C_7-(b-1)\widetilde C_8
=-2(b-1)(b-2)(2b-3).
\label{eq:Eexplicitelimination}
\end{equation}
Since $b-2\neq0$, the only possibilities are
\begin{equation}
(b,\tau)=(1,-2),\qquad
(b,\tau)=\left(\frac32,-2\right).
\label{eq:two-roots}
\end{equation}
Substituting these roots into \eqref{eq:seed} gives the four-dimensional black branes
\begin{equation}
 \dd s_4^2=-F_s(r)\dd v^2+2\dd v\dd r
 +r^2\dd\bm{x}_2^2,\qquad s=a,b,
 \label{eq:4d-metric}
\end{equation}
with
\begin{equation}
\begin{aligned}
 F_a(r)&=\frac{r^2}{L^2}-\mu r,\qquad
 F_b(r)=\frac{r^2}{L^2}-\mu r^{3/2},\\
 \phi(r)&=-\frac2\lambda\log\frac r{r_0},\qquad
 \lambda\chi=-2.
\end{aligned}
 \label{eq:4d-profiles}
\end{equation}
At $b=1$, $\widetilde C_8=0$ fixes the auxiliary coupling combination $v=u_2+16u_1$ to zero; at $b=3/2$, $\widetilde C_7=0$ fixes this combination to $-2$. Using \eqref{eq:Etriangularcouplings} and writing $\zeta=u_1$, we obtain the corresponding coupling vectors
\begin{align}
 E_a:\quad (u_0,u_1,u_2,u_3,u_4)
 &=(6,\zeta,-16\zeta,24\zeta,-32\zeta),
 \label{eq:Ea-4d}\\
 E_b:\quad (u_0,u_1,u_2,u_3,u_4)
 &=(6,\zeta,-2-16\zeta,6+24\zeta,-10-32\zeta).
 \label{eq:Eb-4d}
\end{align}

Each root and its couplings solve all eleven coefficient equations. The determinant calculations and explicit nonzero null vector in Appendix~\ref{app:Ecoefficients} establish that the coefficient matrix has rank four. The original radial equations are also satisfied at arbitrary positive radius:
\begin{equation}
 \E_N=\E_F=\E_\Sigma=\E_\phi=0.
 \label{eq:4d-residuals}
\end{equation}
Thus \eqref{eq:two-roots} exhausts the nondegenerate four-dimensional families within \eqref{eq:seed} with nonzero couplings and a continuous blackening parameter.

\section{\texorpdfstring{Two exact families in $d\geq4$}
{Two exact families in d greater than or equal to 4}}
\label{sec:orbits}

The black-brane field equations determine the coupling orbits as functions of the spacetime dimension $d=p+2$. For a ten-dimensional parent, $n=10-d=8-p$ fixes the internal dimension, and the reduction map is $\mathcal M_{10\to d}(\Theta_{\rm str};n)$. Section~\ref{sec:parent}
matches this map to the black-brane couplings for each $4\leq d\leq9$, the range with a nonempty internal torus.

\subsection{\texorpdfstring{The linear conformally flat orbit $E_a$}
{The linear conformally flat orbit Ea}}

The first root extends to
\begin{equation}
 F_a(r)=\frac{r^2}{L^2}-\mu r,\qquad
 \phi=\chi\log\frac r{r_0},\qquad
 \lambda\chi=-p.
 \label{eq:Ea}
\end{equation}
Let $\zeta\equiv u_1$.  For every $p\geq2$, the eleven coefficient equations are equivalent to four independent coupling conditions, as established in Appendix~\ref{app:Ecoefficients}. They can be written as
\begin{equation}
\begin{gathered}
 u_1=\zeta,\\[-1mm]
 u_0-p(p+1)=0,\qquad
 u_2+4(p-1)(p+2)\zeta=0,\\
 u_3-2p(p-1)(p+4)\zeta=0,\qquad
 u_4+p(p-1)(p^2+5p+2)\zeta=0.
\end{gathered}
 \label{eq:Ea-orbit}
\end{equation}
All five dimensionless couplings are nonzero for $\zeta\neq0$ and $p\geq2$.
Using $\chi=-p/\lambda$ and eliminating $L$ gives the coupling relations for a fixed action
\begin{align}
 p\alpha_0&=(p+1)\lambda^2,\notag\\
 p(p+1)\alpha_2&=-4(p-1)(p+2)\alpha_0\alpha_1,\notag\\
 p^2\alpha_3&=-2(p-1)(p+4)\alpha_1\lambda^3,\notag\\
 p^3\alpha_4&=-(p-1)(p^2+5p+2)\alpha_1\lambda^4.
 \label{eq:Ea-intrinsic}
\end{align}
All eleven radial coefficients and all four original Euler residuals vanish identically for every integer $p\geq2$.

At $p=2$, these relations reproduce \eqref{eq:Ea-4d}. At $p=3$,
\begin{equation}
 \lambda\chi=-3,\qquad
 (u_0,u_1,u_2,u_3,u_4)
 =(12,\zeta,-40\zeta,84\zeta,-156\zeta),
 \label{eq:Ea-5d}
\end{equation}
which gives the five-dimensional member of the family.

The parameter $\zeta=\alpha_1/L^2$ measures the coupling orbit relative to the curvature scale. At fixed couplings and $\lambda$, \eqref{eq:Ea-intrinsic} fixes $\chi=-p/\lambda$ and leaves both
$L$ and $\mu$ free.

\subsection{\texorpdfstring{The fractional Weyl-curved orbit $E_b$}
{The fractional Weyl-curved orbit Eb}}

The second root extends to
\begin{equation}
 F_b(r)=\frac{r^2}{L^2}-\mu r^{3/2},\qquad
 \phi=\chi\log\frac r{r_0},\qquad
 \lambda\chi=-p.
 \label{eq:Eb}
\end{equation}
The eleven coefficient equations again have rank four, and their equivalent independent reduced form is
\begin{equation}
\begin{gathered}
 u_1=\zeta,\\[-1mm]
 u_0-p(p+1)=0,\qquad
 u_2+2+4(p-1)(p+2)\zeta=0,\\
 u_3-3p-2p(p-1)(p+4)\zeta=0,\\
 u_4+p(2p+1)+p(p-1)(p^2+5p+2)\zeta=0.
\end{gathered}
 \label{eq:Eb-orbit}
\end{equation}
The terms proportional to $\zeta$ coincide with $E_a$.  The $\zeta$-independent displacement
\begin{equation}
 (\Delta u_2,\Delta u_3,\Delta u_4)
 =(-2,3p,-p(2p+1))
 \label{eq:affine-displacement}
\end{equation}
distinguishes the fractional solution from $E_a$.  In dimensionful coordinates,
\begin{align}
 \alpha_0&=\frac{p(p+1)}{\chi^2},\notag\\
 \alpha_2&=\frac{-2L^2-4(p-1)(p+2)\alpha_1}{\chi^2},\notag\\
 \alpha_3&=\frac{3pL^2+2p(p-1)(p+4)\alpha_1}{\chi^3},\notag\\
 \alpha_4&=\frac{-p(2p+1)L^2-p(p-1)(p^2+5p+2)\alpha_1}
 {\chi^4}.
 \label{eq:Eb-dimensionful}
\end{align}
Equivalently, the fixed-action equations are
\begin{align}
 p\alpha_0-(p+1)\lambda^2&=0,\notag\\
 2p^2\alpha_3-3p^2\lambda\alpha_2
 -8(p^2-1)\alpha_1\lambda^3&=0,\notag\\
 2p^3\alpha_4-p^2(2p+1)\alpha_2\lambda^2
 -2(p^2-1)(3p+2)\alpha_1\lambda^4&=0,
 \label{eq:Eb-intrinsic}
\end{align}
with
\begin{equation}
 L^2=-\frac12\left[\alpha_2\chi^2
 +4(p-1)(p+2)\alpha_1\right]>0.
 \label{eq:Eb-reality}
\end{equation}
For $p\geq2$, all five dimensionless couplings are nonzero provided
\begin{equation}
 \zeta\notin\left\{0,
 -\frac{1}{2(p-1)(p+2)},
 -\frac{3}{2(p-1)(p+4)},
 -\frac{2p+1}{(p-1)(p^2+5p+2)}\right\}.
 \label{eq:Eb-exclusions}
\end{equation}
All eleven radial coefficients and all four original Euler residuals vanish identically for every integer $p\geq2$.

The displacement \eqref{eq:affine-displacement} fixes $L$ through \eqref{eq:Eb-reality}, while $\mu$ remains a continuous blackening modulus. The positivity condition in \eqref{eq:Eb-reality} selects the AdS region; \eqref{eq:Eb-exclusions} selects its sector with nonzero interaction coefficients.

At $p=2$, these relations reproduce \eqref{eq:Eb-4d}. In five dimensions,
\begin{equation}
 p=3:\quad (u_0,u_1,u_2,u_3,u_4)
 =(12,\zeta,-2-40\zeta,9+84\zeta,-21-156\zeta).
 \label{eq:Eb-anchors}
\end{equation}

\subsection{Two coefficient-selected families}

Both families contain the same five operators but obey distinct fixed-action relations.
At fixed $\lambda$ they share $\chi=-p/\lambda$. Inserting the $E_a$ relation
\begin{equation}
 \alpha_2\chi^2=-4(p-1)(p+2)\alpha_1
\end{equation}
into Eq.~\eqref{eq:Eb-reality} gives $L_b^2=0$. The two sets of coupling relations therefore intersect only at this degenerate limit. No nondegenerate AdS solution belongs to both families. The displacement in Eq.~\eqref{eq:affine-displacement} distinguishes their coupling relations.

Contractions over the transverse directions produce the $p$ dependence of the warped curvature and coefficient equations. The identities in Appendix~\ref{app:Ecoefficients} establish both coupling orbits for every integer $p\geq2$, with their nonzero-coupling domains specified above.

\section{Horizon geometry and entropy}
\label{sec:geometry}

For $E_a$,
\begin{align}
 r_h&=\mu L^2,&T_a&=\frac{\mu}{4\pi}
 =\frac{r_h}{4\pi L^2},\notag\\
 R&=-\frac{(p+1)(p+2)}{L^2}
 +\frac{p(p+1)\mu}{r},&
 C_{\mu\nu\rho\sigma}C^{\mu\nu\rho\sigma}&=0,\notag\\
 X&=\chi^2\left(\frac1{L^2}-\frac\mu r\right),&
 \Box\phi&=\chi\left(\frac{p+1}{L^2}-\frac{p\mu}{r}\right).
 \label{eq:Ea-geometry}
\end{align}
For $E_b$, writing $w=\mu/\sqrt r$,
\begin{align}
 r_h&=\mu^2L^4,&T_b&=\frac{\mu^2L^2}{8\pi}
 =\frac{r_h}{8\pi L^2},\notag\\
 R&=-\frac{(p+1)(p+2)}{L^2}
 +\frac{(2p+1)(2p+3)}4w,&
 C_{\mu\nu\rho\sigma}C^{\mu\nu\rho\sigma}
 &=\frac{p-1}{16(p+1)}w^2,\notag\\
 X&=\chi^2\left(\frac1{L^2}-w\right),&
 \Box\phi&=\chi\left(\frac{p+1}{L^2}-\frac{2p+1}{2}w\right).
 \label{eq:Eb-geometry}
\end{align}
The displayed curvature and scalar invariants are finite at the nonextremal horizons, where $X(r_h)=0$; any divergence occurs only as $r\to0$ for $\mu>0$.  Within the planar ansatz all independent Weyl components are proportional to
\begin{equation}
 {\cal W}[F]=r^2F''-2rF'+2F,
 \qquad {\cal W}[F_a]=0,\qquad
 {\cal W}[F_b]=\frac14\mu r^{3/2}.
 \label{eq:Weylcriterion}
\end{equation}
This proves that $E_a$ is conformally flat and $E_b$ is Weyl curved for every $p\geq2$, in agreement with \eqref{eq:Ea-geometry} and
\eqref{eq:Eb-geometry}.

Both families satisfy the exact measure identity
\begin{equation}
 \sqrt{-g}\,e^{\lambda\phi}
 =r^p\left(\frac r{r_0}\right)^{-p}=r_0^p.
 \label{eq:measure}
\end{equation}
The common exponential cancels the planar area weight pointwise.
The stationary entropy of the complete action inherits this cancellation.

\subsection{Complete stationary Wald entropy}
\label{sec:Wald}

Let $\mathcal B$ be a bifurcation section with induced metric $h_{ij}$, intrinsic scalar curvature $\mathcal R[h]$, tangent projector $h^{\mu\nu}$, and binormal $\epsilon_{\mu\nu}\epsilon^{\mu\nu}=-2$. Write the scalar Lagrangian in \eqref{eq:action}, including $\Omega=e^{\lambda\phi}$ but excluding $1/(16\pi G_d)$, as $\mathcal L_\Omega$. Its complete curvature derivative is
\begin{equation}
\begin{aligned}
 \mathcal P^{\mu\nu\rho\sigma}
 &\equiv\frac{\partial\mathcal L_\Omega}
 {\partial R_{\mu\nu\rho\sigma}}
 =\Omega\left[
 g^{\mu[\rho}g^{\sigma]\nu}
 +\alpha_1\Pi^{\mu\nu\rho\sigma}
 +\alpha_2\mathcal D^{\mu\nu\rho\sigma}\right],\\
 \mathcal D^{\mu\nu\rho\sigma}
 &=\frac14\left(
 g^{\mu\rho}\nabla^\nu\phi \nabla^\sigma\phi-g^{\mu\sigma}\nabla^\nu\phi \nabla^\rho\phi\right.\\
 &\qquad\left.
 -g^{\nu\rho}\nabla^\mu\phi \nabla^\sigma\phi+g^{\nu\sigma}\nabla^\mu\phi \nabla^\rho\phi
 \right)-\frac X2g^{\mu[\rho}g^{\sigma]\nu}.
\end{aligned}
\label{eq:WaldP}
\end{equation}
Here $\Pi^{\mu\nu\rho\sigma}$ is given in \eqref{eq:covarianttensors}; the derivative holds the metric and scalar covariant derivatives fixed and respects the Riemann symmetries.
The $\alpha_0X$ and $\alpha_4X^2$ terms contain no curvature. The cubic scalar interaction contributes only a term proportional to the Killing vector in the bifurcation charge, and hence no entropy; its explicit reduction is given in Appendix~\ref{app:Wald}.

With unit-weight antisymmetrization, the three binormal contractions are
\begin{equation}
\begin{gathered}
 g^{\mu[\rho}g^{\sigma]\nu}
 \epsilon_{\mu\nu}\epsilon_{\rho\sigma}=-2,\qquad
 \Pi^{\mu\nu\rho\sigma}\epsilon_{\mu\nu}\epsilon_{\rho\sigma}
 =-4\mathcal R[h],\\
 \mathcal D^{\mu\nu\rho\sigma}
 \epsilon_{\mu\nu}\epsilon_{\rho\sigma}
 =h^{\mu\nu}\nabla_\mu\phi\nabla_\nu\phi
 \equiv(D\phi)^2.
\end{gathered}
\label{eq:Waldcontractions}
\end{equation}
In the last identity, the normal-gradient contributions from $R_{\mu\nu}\nabla^\mu\phi \nabla^\nu\phi$ and $-RX/2$ cancel. The Gauss equation in the second identity uses the vanishing extrinsic curvatures of a smooth bifurcation section \cite{JacobsonMyers1993}. The stationary Wald entropy is therefore
\cite{Wald1993,IyerWald1994}
\begin{equation}
\begin{aligned}
 S_{\rm W}
 &=-\frac1{8G_d}\int_{\mathcal B}\dd^px\,\sqrt h\,
 \mathcal P^{\mu\nu\rho\sigma}
 \epsilon_{\mu\nu}\epsilon_{\rho\sigma}\\
 &=\frac1{4G_d}\int_{\mathcal B}\dd^px\,\sqrt h\,\Omega
 \left[1+2\alpha_1\mathcal R[h]
 -\frac{\alpha_2}{2}(D\phi)^2\right].
\end{aligned}
\label{eq:Waldfunctional}
\end{equation}

Both families extend smoothly through their bifurcation surfaces. Near their positive-radius horizons, $F=2\kappa(r-r_h)+O((r-r_h)^2)$ with $\kappa>0$.
In regular Euclidean polar coordinates, $r-r_h=\kappa\rho^2/2+O(\rho^4)$, and
\begin{equation}
 \phi=\phi_h+\frac{\chi\kappa}{2r_h}\rho^2+O(\rho^4).
 \label{eq:Waldsmooth}
\end{equation}
With Euclidean time period $2\pi/\kappa$, both the metric and the scalar are smooth at the origin and admit a smooth local Kruskal extension to $\mathcal B$.

For the flat sections and radial scalars considered here, $\mathcal R[h]=0$ and $D_i\phi=0$. Thus the Gauss--Bonnet and Einstein-tensor derivative couplings give no entropy correction.
Writing $V_p$ for the regulated coordinate volume of the plane, the complete entropy density of either family is
\begin{equation}
 s_{{\rm W},a}=s_{{\rm W},b}
 =\frac{S_{\rm W}}{V_p}
 =\frac{r_h^p e^{\lambda\phi_h}}{4G_d}
 =\frac{r_0^p}{4G_d},\qquad d=p+2\geq4.
 \label{eq:Waldentropy}
\end{equation}
For $G_d>0$ and $r_0>0$ this is positive and independent of the blackening modulus at fixed $r_0$. It is also independent of $L$, although the two families have different temperatures and belong to different coupling loci.

At fixed couplings, $E_a$ fixes $\chi=-p/\lambda$ but leaves both $L$ and $\mu$ free. For $E_b$, Eq.~\eqref{eq:Eb-reality} fixes $L$, while $\mu$ remains free. Both branches therefore contain continuous families of nonextremal horizons with $\delta_\mu S_{\rm W}=0$ at fixed $r_0,G_d,p,V_p$.
Variation of the scalar zero mode gives $\delta S_{\rm W}=pS_{\rm W}\delta r_0/r_0$ at fixed $G_d,p,V_p$.

\subsection{Entropy mechanism and Einstein variables}
\label{sec:EinsteinEntropy}

The constant entropy has a local geometric mechanism. Along a variation of the horizon radius at fixed scalar zero mode,
$\delta\log\sqrt h=p\,\delta r_h/r_h$ and $\delta\log\Omega_h=\lambda\chi\,\delta r_h/r_h =-p\,\delta r_h/r_h$. The intrinsic-curvature and tangential-gradient corrections in \eqref{eq:Waldfunctional} vanish independently, so the cancellation holds for the complete entropy while all derivative interactions continue to support the field equations.

Einstein variables give the entropy cancellation a direct geometric interpretation. In $d=p+2$, the transformation
\begin{equation}
\begin{aligned}
 g^E_{\mu\nu}&=\Omega^{2/p}g_{\mu\nu}
 =\left(\frac{r_0}{r}\right)^2g_{\mu\nu},\qquad
 A_E=V_pr_0^p=4G_dS_{\rm W},\\
 \alpha_0^{E,(2)}&=\alpha_0-\frac{p+1}{p}\lambda^2=0
\end{aligned}
\label{eq:Einstein-area}
\end{equation}
puts the two-derivative curvature term in Einstein form. The transverse Einstein-frame area is independent of the horizon radius. The complete entropy calculated above therefore equals this constant area divided by $4G_d$. The same coupling relation cancels the explicit two-derivative scalar kinetic term after the conformal transformation and a covariant integration by parts. The transformed action retains the higher-derivative scalar interactions. The conformal factor is finite and positive at either horizon and tends to zero at infinity.

This gives a concrete realization of black holes whose horizon radius and temperature vary while their complete stationary entropy remains fixed. Constant entropy along a black-hole family also occurs in the conformal-gravity construction of Ref.~\cite{FanLu2015}, in a sector with a ghost-like Maxwell field. Here the mechanism is the scalar weight in the complete Lovelock--Horndeski entropy functional.

\section{Inverse compactification mapping and a no-go theorem}
\label{sec:parent}

Matching the black-brane coupling loci to a higher-dimensional reduction defines an algebraic inverse compactification problem. For a real one-scalar flat-torus reduction, the two-derivative kinetic relation gives an obstruction independent of the four-derivative frame coefficients.

Consider the two-derivative sector of a $(d+n)$-dimensional metric--dilaton parent,
\begin{equation}
 \widehat S^{(2)}=\frac{1}{16\pi G_{d+n}}
 \int\dd^{d+n}x\sqrt{-\widehat g}\,e^{-2\widehat\Phi}
 \left[\widehat R+4(\widehat\nabla\widehat\Phi)^2\right],
 \label{eq:parent-action}
\end{equation}
The curvature-squared and four-derivative parent sectors carry the same $e^{-2\widehat\Phi}$ prefactor. We use the real one-scalar flat-torus ansatz
\begin{equation}
 \dd\widehat s_{d+n}^2=e^{2\alpha_{\rm K}\phi}\dd s_d^2
 +e^{2\beta_{\rm K}\phi}\dd\bm{y}_n^2,\qquad
 \widehat\Phi=s\phi .
 \label{eq:parent-ansatz}
\end{equation}
For the flat internal space, the two- and four-derivative sectors acquire the exponential weights
\begin{equation}
 \Omega_{(2)}=e^{[(d-2)\alpha_{\rm K}+n\beta_{\rm K}-2s]\phi},
 \qquad
 \Omega_{(4)}=e^{[(d-4)\alpha_{\rm K}+n\beta_{\rm K}-2s]\phi}.
 \label{eq:parent-sector-weights}
\end{equation}
Matching both to the common factor $e^{\lambda\phi}$ gives $\alpha_{\rm K}=0$ and $n\beta_{\rm K}=\lambda+2s$.  With these choices,
the warped-product Ricci scalar and dilaton kinetic term are
\begin{equation}
 \widehat R=R-2n\beta_{\rm K}\Box\phi
 -n(n+1)\beta_{\rm K}^2X,\qquad
 4(\widehat\nabla\widehat\Phi)^2=4s^2X.
 \label{eq:parent-warped-kinetics}
\end{equation}
The constant coordinate volume of the internal torus is absorbed into the lower-dimensional Newton constant.  Integrating the Laplacian with its scalar weight gives
\begin{equation}
 \int\dd^dx\sqrt{-g}\,e^{\lambda\phi}\Box\phi
 =-\lambda\int\dd^dx\sqrt{-g}\,e^{\lambda\phi}X
 +\text{boundary term}.
 \label{eq:parent-weighted-ibp}
\end{equation}
Consequently the coefficient of $X$ inside the weighted reduced Lagrangian is
\begin{equation}
 \alpha_0^{\rm KK}
 =4s^2-n(n+1)\beta_{\rm K}^2+2n\beta_{\rm K}\lambda.
 \label{eq:parent-kinetic-intermediate}
\end{equation}
Substituting $n\beta_{\rm K}=\lambda+2s$ gives the coefficient relations
\cite{VanAcoleyenVanDoorsselaere2011,
CharmousisGouterauxKiritsis2012,WuStone2026}
\begin{equation}
 \alpha_{\rm K}=0,\qquad n\beta_{\rm K}=\lambda+2s,
 \qquad
 \alpha_0^{\rm KK}
 =\frac{(n-1)\lambda^2-4s\lambda-4s^2}{n}.
 \label{eq:parent-kinetic}
\end{equation}
Both exact E orbits require
\begin{equation}
 \alpha_0^E=\frac{p+1}{p}\lambda^2.
 \label{eq:E-kinetic}
\end{equation}
Their intersection is governed by
\begin{equation}
 (p+n)\lambda^2+4ps\lambda+4ps^2=0,
 \qquad \Delta_\lambda=-16pns^2\leq0.
 \label{eq:obstruction}
\end{equation}
For $p,n>0$ and $s\neq0$, the discriminant is strictly negative. Completing the square treats all real $s$ uniformly:
\begin{equation}
 (p+n)\left(\lambda+\frac{2ps}{p+n}\right)^2
 +\frac{4pns^2}{p+n}=0.
 \label{eq:parent-positive-form}
\end{equation}
Both terms are nonnegative for real $\lambda,s$.  If $s\neq0$, the second is positive; if $s=0$, the equation requires $\lambda=0$, which is excluded by $\lambda\chi=-p$ on either black-brane family.  The specified one-scalar flat-torus reduction therefore cannot reproduce the couplings of either family for any real $s$.  The obstruction already occurs in the two-derivative kinetic coefficient and is independent of the remaining four-derivative frame parameters.

For a ten-dimensional parent, $n=8-p$ and $p=2,\ldots,7$.
The dimension-dependent two-derivative map becomes
\begin{equation}
\beta_{\rm K}^{(p)}=\frac{\lambda+2s}{8-p},\qquad
\alpha_0^{\rm KK}(p)
=\frac{(7-p)\lambda^2-4s\lambda-4s^2}{8-p}.
\label{eq:KKtenexplicit}
\end{equation}
Consequently, the intersection condition is
\begin{equation}
8\left(\lambda+\frac{ps}{4}\right)^2
+\frac{p(8-p)}2s^2=0.
\label{eq:KKtenpositive}
\end{equation}
This comparison uses the appropriate internal dimension at every $d$, and gives the same obstruction throughout the physical range.
The ten-to-four and ten-to-five examples are

\begin{equation}
 10\to4:\quad \lambda^2+\lambda+1=0,
 \qquad
 10\to5:\quad 2\lambda^2+3\lambda+3=0
 \label{eq:anchors}
\end{equation}
for $s=1$. Curved internal geometry, fluxes and independent breathing--dilaton fields modify the coefficient map defined by
\eqref{eq:parent-ansatz}.

The same separation is visible in a ratio invariant under scalar normalization.  The standard flat neutral Lovelock--Galileon reduction carries
\begin{equation}
 \left(\frac{\alpha_0}{\lambda^2}\right)_{\rm GL}
 =\frac{n-1}{n},
 \qquad
 \left(\frac{\alpha_0}{\lambda^2}\right)_{E_a,E_b}
 =\frac{p+1}{p}.
 \label{eq:kinetic-invariant}
\end{equation}
Equality requires the formal continuation $n=-p$. Thus the two E orbits and the conventional positive-dimensional flat reduction have different real kinetic ratios \cite{CharmousisGouterauxKiritsis2012}.

The kinetic relation \eqref{eq:E-kinetic} controls both the Einstein-variable cancellation in \eqref{eq:Einstein-area} and the compactification obstruction. The nonlinear horizon equations thus distinguish theories with the same curvature--scalar operator content through their coupling correlations. This links the constant-entropy mechanism to the compactification data of a higher-dimensional parent.

\section{Conclusions}
\label{sec:discussion}
\label{sec:conclusion}

The two black-brane families reveal a direct relation between scalar hair, horizon entropy and the higher-dimensional origin of Lovelock--Horndeski gravity. Their radii and temperatures vary along continuous families while the complete stationary entropy remains constant when the scalar zero mode is held fixed. The scalar weight compensates the geometric area, and the Einstein variables express this mechanism as a constant transverse horizon area. This behavior follows from the full action and accompanies a sharp distinction between the two geometries: the linear family is conformally flat with a free AdS scale, while the fractional family is Weyl curved with its scale fixed by the couplings. The four-dimensional classification determines both branches in the nondegenerate sector with a logarithmic scalar and a single-power metric deformation. Their explicit continuation to every $d\geq4$ shows how the nonlinear field equations sustain these horizons. The common kinetic relation also leads to the inverse compactification no-go: the specified real one-scalar flat-torus reduction cannot reproduce either coupling family,
independently of its four-derivative frame parameters. The heterotic construction \cite{WuStone2026} motivates the interaction structure, while the obstruction identifies coupling relations that require a different higher-dimensional realization. These exact backgrounds provide a setting for extending the boundary-response construction of ~\cite{WuHolographicRenormalization2026} to dilaton-coupled scalar--tensor gravity.

\appendix
\section{Coefficient equations and their rank}
\label{app:Ecoefficients}

\subsection{Four-dimensional coefficient equations}

Use the coefficient coordinates \eqref{eq:ucoordinates}, equivalently $L=\chi=1$, and remove the nonzero weight and volume from each Euler expression:
\begin{equation}
\varepsilon_Q=r^{-p}e^{-\lambda\phi}\E_Q,\qquad
R_{Qj}=[\mu^j]\varepsilon_Q\big|_{r=1},\qquad
Q=N,F,\Sigma,\phi.
\label{eq:Egeneratorprovenance}
\end{equation}
The scalar zero-mode factor cancels in this definition. The generator normalizations are fixed by
\begin{equation}
\begin{aligned}
(C_1,C_2,C_3)&=(R_{N0},-R_{N1},R_{N2}),&
(C_4,C_5)&=(R_{F0},-R_{F1}),\\
(C_6,C_7,C_8)&=(\tfrac12R_{\Sigma0},-R_{\Sigma1},R_{\Sigma2}),&
(C_9,C_{10},C_{11})&=(-R_{\phi0},R_{\phi1},R_{\phi2}).
\end{aligned}
\label{eq:Enormalizations}
\end{equation}
There is no quadratic $\mu$ coefficient in the $F$ equation.
Specializing only now to $p=2$ yields
\begin{align}
C_1={}&u_0-2\tau^2-4\tau-6+8\tau(\tau+2)u_1
-(2\tau+3)u_2-\tau u_3+u_4,\notag\\
C_2={}&u_0-(\tau+2)b-2\tau^2-2\tau-2
+4\tau(3b+4\tau+2)u_1\notag\\
&-(3b+4\tau)u_2-(b+2\tau-2)u_3+2u_4,\notag\\
C_3={}&4\tau(3b+2\tau-2)u_1-(3b+2\tau-3)u_2
-(b+\tau-2)u_3+u_4,
\label{eq:Egenn}\\
C_4={}&u_0-\tau^2+\tau+4\tau(\tau-1)u_1
+(3-\tau)(u_2+u_3)+2u_4,\notag\\
C_5={}&4\tau(\tau-1)u_1+(3-\tau)(u_2+u_3)+2u_4,
\label{eq:Egenf}\\
C_6={}&C_1,\notag\\
C_7={}&2u_0-2b^2-4b\tau-2b-4\tau^2
+8\tau(b^2+b\tau+2b+2\tau)u_1\notag\\
&-(b^2+b\tau+5b+6\tau-2)u_2
-2(b+2\tau-2)u_3+4u_4,\notag\\
C_8={}&8\tau b(2b+\tau-2)u_1
-(2b^2+b\tau+b+2\tau-4)u_2
-2(b+\tau-2)u_3+2u_4,
\label{eq:EgenSigma}\\
C_9={}&(\tau+6)u_0-24\tau u_1+3(\tau+6)u_2
+(18-\tau^2)u_3+3(\tau+4)u_4+12\tau,\notag\\
C_{10}={}&(2b+\tau+2)u_0-4\tau(b^2+3b+2)u_1
+(2b^2+b\tau+10b+4\tau+8)u_2\notag\\
&+(b^2-2b\tau+11b-2\tau^2+4\tau+10)u_3
+(8b+6\tau+8)u_4+\tau(b^2+3b+2),\notag\\
C_{11}={}&4\tau b(2b-1)u_1-(4b^2+b\tau+2b+\tau-2)u_2\notag\\
&+(-2b^2+2b\tau-7b+\tau^2-4\tau+4)u_3
-(8b+3\tau-4)u_4.
\label{eq:Egenphi}
\end{align}
The duplicate $C_6=C_1$ is retained to show its origin in the transverse equation. All eleven rows are affine in the five $u_i$.
Equations~\eqref{eq:ideal-factors}--\eqref{eq:Eexplicitelimination}
are polynomial identities between these displayed generators, followed
by elimination in the physical domain $b<2$.

\subsection{Rank of the coupling matrix in arbitrary dimension}

For general $p$, define $C_i(p)$ by the same extraction and the same normalizations \eqref{eq:Egeneratorprovenance} and \eqref{eq:Enormalizations}, before setting $p=2$. On $\tau=-p$, consider their coefficient matrix in the four columns $(u_0,u_2,u_3,u_4)$. For both exponents, the first three selected rows $(C_4,C_1,C_5)$ form the common block
\begin{equation}
B(p)=\begin{pmatrix}
1&\tfrac12p(2p+1)&2p+1&2\\
1&\tfrac12p(p-1)&p&1\\
0&\tfrac12p(2p+1)&2p+1&2
\end{pmatrix}.
\label{eq:Erankcommon}
\end{equation}
For $E_a$, append row $C_8$; for $E_b$, append row $C_7$. The resulting
matrices and their exact determinants are
\begin{equation}
\begin{aligned}
M_a(p)&=\begin{pmatrix}
\multicolumn{4}{c}{B(p)}\\[1mm]
0&\tfrac12p(p^2+2p-1)&p(p+1)&p
\end{pmatrix},&\det M_a&=\frac p2,\\
M_b(p)&=\begin{pmatrix}
\multicolumn{4}{c}{B(p)}\\[1mm]
p&\tfrac18p(8p^2-2p+1)&\tfrac12p(4p+1)&2p
\end{pmatrix},&\det M_b&=-\frac p8.
\end{aligned}
\label{eq:Erankminors}
\end{equation}
The determinants are nonzero throughout $p\geq2$. The full \(11 \times 5\) coefficient matrix therefore has rank at least four.
With columns ordered as $(u_0,u_1,u_2,u_3,u_4)$, its null vector is
\begin{equation}
\boldsymbol v_p=\left(
0,1,-4(p-1)(p+2),2p(p-1)(p+4),
-p(p-1)(p^2+5p+2)\right).
\label{eq:Eranknull}
\end{equation}
Substituting \eqref{eq:Ea-orbit} or \eqref{eq:Eb-orbit}, with free $u_1=\zeta$, annihilates every coefficient identically in $p$. Differentiating these identities with respect to $\zeta$ verifies the null vector and establishes the upper bound four. Thus the coefficient matrix has rank exactly four for every integer $p\geq2$. At $p=2$, the determinants evaluate to $\det M_a=1$ and $\det M_b=-1/4$.

Finally, the same substitutions in the uncollected radial expressions give, for $s=a,b$,
\begin{equation}
(C_1^{(s)}(p),\ldots,C_{11}^{(s)}(p))=(0,\ldots,0),
\label{eq:elevenrowresiduals}
\end{equation}
\begin{equation}
(\E_N,\E_F,\E_\Sigma,\E_\phi)=(0,0,0,0).
\label{eq:foureulerresiduals}
\end{equation}
These identities hold for arbitrary $r>0$ and $\mu$, and for every integer $p\geq2$.
For $E_b$, powers are interpreted on $r=x^2$, $x>0$.
The four-dimensional completeness statement follows from the triangular elimination; the general-$p$ identities establish the continuation of both families to every \(d \ge 4\).

\section{Binormal contraction and the cubic scalar term}
\label{app:Wald}

The contractions in \eqref{eq:Waldcontractions} hold off shell. In an orthonormal normal plane, take $\epsilon_{01}=1$, and let $g^\perp_{\mu\nu}=g_{\mu\nu}-h_{\mu\nu}$.
The Riemann derivative of $R_{\mu\nu}\nabla^\mu\phi \nabla^\nu\phi$ contracts to $-g^\perp_{\mu\nu}\nabla^\mu\phi \nabla^\nu\phi$, while that of $-RX/2$ contracts to $X$.
Their sum is precisely $(D\phi)^2$.

There is also a direct coordinate derivation for arbitrary $F(r)$ and a radial scalar in the metric
\[
 \dd s^2=-F\dd t^2+\frac{\dd r^2}{F}+r^2\dd\bm{x}_p^2.
\]
Taking $\epsilon_{tr}=1$ in the exterior gives
\begin{equation}
 g^{t[t}g^{r]r}=-\frac12,\qquad
 \Pi^{trtr}=\frac{p(p-1)F}{r^2},\qquad
 \mathcal D^{trtr}=-\frac{F(\phi')^2}{4}+\frac X4=0.
 \label{eq:Waldcomponents}
\end{equation}
Consequently
\begin{equation}
 \mathcal P^{\mu\nu\rho\sigma}
 \epsilon_{\mu\nu}\epsilon_{\rho\sigma}
 =4\mathcal P^{trtr}
 =\Omega\left[-2+\frac{4\alpha_1p(p-1)F}{r^2}\right].
 \label{eq:Waldcoordinate}
\end{equation}
The binormal contraction approaches $-2\Omega_h$ at the horizon and extends smoothly to the bifurcation surface through \eqref{eq:Waldsmooth}.

For the cubic interaction, set $A_3=\alpha_3\Omega X$ and vary $\mathcal L_3=A_3\Box\phi$. With $\delta g_{\mu\nu}=\gamma_{\mu\nu}$ and $\gamma=g^{\mu\nu}\gamma_{\mu\nu}$, its symplectic-potential vector, with the common action normalization suppressed, is
\begin{equation}
\begin{aligned}
 \Theta_3^\mu
 ={}&-A_3\nabla_\nu\phi\gamma^{\mu\nu}
 +\frac12A_3\nabla^\mu\phi\gamma
 +A_3\nabla^\mu\delta\phi\\
 &+(2\alpha_3\Omega(\Box\phi)\nabla^\mu\phi-\nabla^\mu A_3)\delta\phi .
\end{aligned}
\label{eq:cubicTheta}
\end{equation}
Here $\gamma^{\mu\nu}=g^{\mu\rho}g^{\nu\sigma}\gamma_{\rho\sigma}
=-\delta g^{\mu\nu}$.
On substituting
$\gamma_{\mu\nu}=2\nabla_{(\mu}\xi_{\nu)}$ and $\delta\phi=\xi^\mu \nabla_\mu\phi$, the derivative-of-$\xi$ terms reduce to $A_3(\nabla^\mu\phi\nabla_\nu\xi^\nu-\nabla^\nu\phi\nabla_\nu\xi^\mu)$.
Extracting a divergence yields an antisymmetric potential proportional to $A_3(\nabla^\mu\phi\xi^\nu-\nabla^\nu\phi\xi^\mu)$, with no $\nabla_\mu\xi_\nu$ term. It vanishes at $\mathcal B$, where $\xi^\mu=0$. The cubic interaction therefore contributes zero bifurcation entropy.

The same conclusion follows after a regular covariant integration by parts of the Lagrangian:
\begin{equation}
 \Omega X\Box\phi
 =\nabla_\mu(\Omega X\nabla^\mu\phi)
 -\Omega\left(2\nabla^\mu\phi\nabla^\nu\phi\,\nabla_\mu\nabla_\nu\phi+\lambda X^2\right).
 \label{eq:cubicIBP}
\end{equation}
The resulting charge shift is proportional to the contraction of $\xi$ with the boundary form, so it vanishes on the smooth bifurcation
section \cite{IyerWald1994,JacobsonKangMyers1994}. The complete bifurcation entropy is therefore \eqref{eq:Waldentropy}.

\begingroup
\raggedright

\endgroup

\end{document}